# Breakdown of the Stokes-Einstein Relation Above the Melting Temperature in a Liquid Phase-Change Material

Shuai Wei[1], Zach Evenson[2*], Moritz Stolpe[3], Pierre Lucas[4], C. Austen Angell[5*]

[1]*I. Institute of Physics(IA), RWTH Aachen University, Aachen, Germany*

[2]*Maier-Leibnitz Zentrum (MLZ) and Physik Department, Technische Universität München, Lichtenbergstrasse 1, 85747 Garching, Germany*

[3]*Heraeus Holding GmbH, Germany*

[4]*Department of Materials Science and Engineering, University of Arizona, Tucson, Arizona 85712, United States.*

[5]*School of Molecular Sciences, Arizona State University, Tempe, Arizona 85287, United States.*

## ABSTRACT

The dynamic properties of liquid phase-change materials (PCMs), such as viscosity $\eta$ and atomic self-diffusion coefficients $D$, play an essential role in ultrafast phase switching behavior of novel non-volatile phase-change memory applications, as they are intimately related to crystallization kinetics and phase stabilities. To connect $\eta$ to $D$, the Stokes-Einstein relation (SER) is commonly assumed to be valid at high temperatures near or above the melting temperature $T_m$ and is frequently employed for assessing liquid fragility (or crystal growth velocity) of technologically important PCM compositions. However, using quasi-elastic neutron scattering (QENS), we give here experimental evidence for a breakdown of the SER even at temperatures above $T_m$ in the high-atomic-mobility state of a typical PCM, $Ge_1Sb_2Te_4$, where the decay of density correlation functions still remains exponential. The origin of the breakdown is thus unlikely the result of dynamical heterogeneities, as is usually postulated for viscous liquids. Rather, we discuss its possible connections to a metal-semiconductor and fragile-strong transition hidden below $T_m$.

## INTRODUCTION

PCMs can be reversibly switched between their glassy and crystalline states by heating with a voltage or laser pulse[1]. The strong optical/electrical contrast between these two states makes PCMs highly interesting for data storage applications (e.g. encoding "0" and "1"). An extremely fast phase switching on a time-scale of nanoseconds is a requirement for fast read/write speed. However, the fast atomic dynamics inherent to PCMs seems to be at odds with the concomitant requirement of good amorphous phase stability for data retention[1]. Typical PCMs include the Ge-Sb-Te alloys, especially those along the GeTe - $Sb_2Te_3$ tie line, and doped $Sb_2Te$ alloys such

---

as Ag-In-Sb-Te(*2*). Physical understanding of PCMs has been mainly centered around features of their crystalline states (e.g. bonding(*3*) and vacancy ordering(*4*)), while the liquid-state behavior was considered 'ordinary' or less explored, probably because a large portion of the (supercooled) liquid state is obscured by fast crystallization.

Requiring a critical cooling rate of $\sim 10^{10}$ K s$^{-1}$ for vitrification, PCMs are generally recognized as poor glass formers. In fact, the amorphous phase is so prone to crystallization that no glass transition ($T_g$) can be observed in a differential scanning calorimeter (DSC)(*5*) before crystallization sets in upon heating(*6*). Thus, the broad supercooled liquid regime, $\Delta T = T_m - T_x$, between the melting temperature (e.g. $T_m \sim 903$ K for Ge$_1$Sb$_2$Te$_4$) and the crystallization temperature upon heating (typically $T_x \sim 400$ K for GeTe - Sb$_2$Te$_3$ alloys) is experimentally inaccessible using standard techniques. For this reason, it has been a long-standing challenge to characterize the liquid-state behavior of PCMs -- specifically, the liquid fragility that has been recently given much importance by Greer and coworkers(*7*). The latter, defined as $m = dlog\eta/d(T_g/T)|_{T=Tg}$, where $T_g$ is the "standard" value (where the viscosity $\eta$ reaches the value $10^{12}$ Pa·s)(*8*), describes the deviation of the temperature-dependence of viscosity from the Arrhenius law. Fragility has been recognized as a useful parameter for understanding crystallization kinetics and the stability of amorphous states(*9, 10*).

The SER is frequently employed to calculate $\eta$ from $D$ (or vice-versa) in technologically important PCMs at 'sufficiently' high temperature,

$$D \cdot \eta = (k_B \cdot T)/(6\pi r_H), \qquad (1)$$

where $k_B$ is the Boltzmann constant, $T$ the absolute temperature, and $r_H$ the effective hydrodynamic radius. For instance, Orava et al.(*7*) assumed a valid SER at $T_m$ for deriving the absolute values of crystal growth velocity of Ge$_2$Sb$_2$Te$_5$ from the Kissinger-type analysis using ultrafast DSC data. Salinga et al.(*11*) used crystal growth velocity data from laser reflectivity measurements for determining the fragility of Ag-In-Sb-Te ($m \sim 190$), assuming a valid SER over a wide temperature range well below $T_m$. Schumacher et al.(*12*) compared the experimental viscosity data for Ge$_2$Sb$_2$Te$_5$ at $T > T_m$ to the values derived from the SER based on simulated self-diffusion coefficients, and observed a non-negligible discrepancy.

In general, liquids at high temperature are expected to obey the SER, as they do not feel the energy landscape and single-particle dynamics follow the same temperature-dependence as the collective macroscopic stress relaxation processes. When the temperature approaches $T_g$ on cooling, $D$ progressively decouples from $\eta$ in fragile liquids such as o-terphenyl (OTP), PDE, and salol, beginning at $\sim 1.2 T_g$ (ref.[13]), supposedly due to the dynamic heterogeneity[13]. A SER breakdown in PCM GeTe was also asserted by Sosso et al.[14] based on ab initio simulations, which occurs in the supercooled liquid attributed to dynamic heterogeneities. Such a breakdown in supercooled $Ge_2Sb_2Te_5$ and GeTe nanoparticles was also taken into account by Chen et al. [15] and Orava et al.[7], where the necessity of using a fractional SER to describe the supercooled liquid, was emphasized. Also, the crystal growth kinetic coefficient $U_{kin}$ decouples from $\eta$ following the Ediger et al relation[9] $U_{kin} \propto \eta^{-\xi^*}$ ( $\xi^* < 1$ depending on fragility). Detailed experimental studies of multi-component bulk metallic glass-forming liquids with high atomic packing fractions[16] ($\varphi \sim 0.51$-0.55) revealed a clear breakdown of the SER close to, or even well above, the critical temperature $T_c$ of mode-coupling theory (MCT)[17–19]. Liquid PCMs, on the other hand, represent $p$-electron bonded[20, 21] fragile glass formers[7, 22] with low atomic packing fractions ($\varphi \sim 0.3$-0.4). There has been surprisingly little interest in, or experimental data related to, the SER above $T_m$ for PCMs. However, the recent discussion concerning the likely existence of liquid-liquid transitions (LLTs) in PCMs suggests the probability of complex dynamical behavior, including a breakdown of the SER in these liquids[23].

In this work, we probe the microscopic dynamics in the liquid state of a typical PCM $Ge_1Sb_2Te_4$ using QENS, which permits direct determination of both the structural $\alpha$-relaxation time, (proportional to shear viscosity $\eta$), and the self-diffusion coefficient $D$ on the same sample under identical conditions. Our results question the validity of the commonly employed SER in those technologically important PCMs, even well above $T_m$. We discuss the origin of the breakdown of SER and its relation to a possible metal-semiconductor and fragile-strong transition hidden below $T_m$, which may play a critical role in speeding up crystallization kinetics, before restraining the atomic rearrangements through a fragile-to-strong transition. The fundamental importance of such phenomena to the technical performance of PCMs has been stressed elsewhere[23, 24].

## RESULTS

### α-relaxation time

We obtain relaxation times from the decay of the intermediate scattering function (ISF) $S(q,t)$ which describes the decay of microscopic density fluctuations in the liquid and was obtained according to the procedure outlined in the Materials and Methods. Figure 1a shows $S(q,t)$ taken at the first structure factor maximum $q_0$=2.0 Å$^{-1}$ of the liquid at different temperatures above $T_m$= 903

K. The data are best fitted with a simple exponential function, $S(q,t)/S(q,0) = f_q \, exp(-t/\tau_q)$, where $f_q$ is a constant accounting for atomic vibrations and $\tau_q$ is the structural relaxation time. In the case of $q_0$=2.0 A$^{-1}$, i.e. the position of the structure factor maximum, the fitting yields a collective structural relaxation time, or $\alpha$-relaxation time $\tau_\alpha$, shown in Figure 1b, as the quasi-elastic signal at $q_0$ arises predominantly from the coherent scattering contribution. $\tau_\alpha$ is associated with the shear viscosity $\eta$ in the viscoelastic model of Maxwell, which establishes a proportional relation via $\eta = G_\infty \cdot \tau_\alpha$, where $G_\infty$ is the infinite-frequency shear modulus measured on time scales very short with respect to $\tau_\alpha$. This proportionality has been directly tested by combining QENS and viscosity measurements on various glass forming melts([18, 25, 26]).

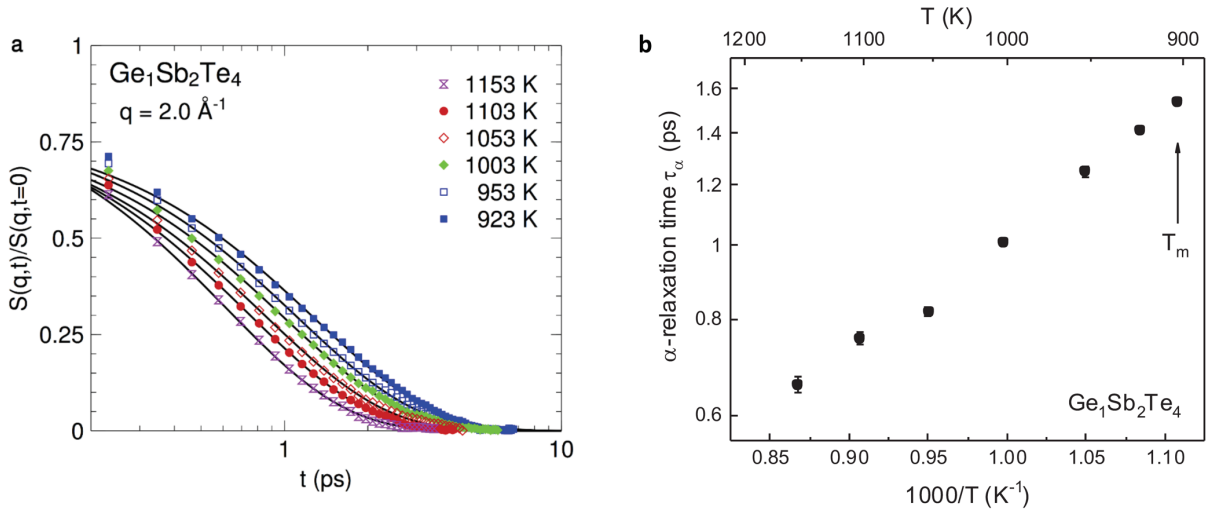

**Figure 1. Exponential decay of the ISF and the $\alpha$-relaxation time $\tau_\alpha$ for liquid Ge$_1$Sb$_2$Te$_4$ above $T_m$.** (**a**) The decay of the ISF $S(q,t)$ of liquid Ge$_1$Sb$_2$Te$_4$ at the structure factor first maximum $q_0$=2.0 Å$^{-1}$ measured at temperatures above $T_m$=903 K. The data, typical of high fluidity systems, are best fitted by simple exponential functions (solid lines), each yielding a single relaxation time $\tau_\alpha$ (see text for details). Note that at very short time (<0.65 ps) the data points correspond to phonons and fast processes that are not explicitly taken into account in the fitting. This is consistent with the analysis of the dynamic structure factor $S(q, \omega)$ in the energy transfer ($\hbar\omega$) domain (see SI-Fig.S1), where $S(q, \omega)$ is best described by a single Lorentzian form. (**b**) Arrhenius plot for the $\alpha$-relaxation time $\tau_\alpha$ above $T_m$.

## Self-diffusivity

Self-diffusion coefficients were determined from the QENS signal in the low-$q$ range, where the signal is dominated by the incoherent scattering of both Ge and Te atoms and reflects their single-particle dynamics on long length and time scales. Given the incoherent cross-sections of each species and their relative concentration in the alloy melt, the measured self-diffusion coefficient of $Ge_1Sb_2Te_4$ represents a mean value weighted by roughly 1/3 Ge and 2/3 Te. As shown in the inset of Fig. 2, the incoherent relaxation times $\tau^{inc}$ indeed follow a $1/q^2$ dependence at low $q^2 \leq 0.6$ Å$^{-2}$, which is characteristic of long-range atomic diffusion in liquids in the hydrodynamic limit as $q \rightarrow 0$ Å$^{-1}$ (ref.(27)). This thus allows us to derive a mean Ge/Te self-diffusion coefficient via $D_{Ge/Te} = 1/(\tau^{inc}q^2)$. In Figure 2, the resulting $D_{Ge/Te}$ are fitted with the Arrhenius law, yielding an activation energy $E_{a,D}$ = 26.41±0.89 kJ mol$^{-1}$ and a pre-exponent $D_0$ = 1.4×10$^{-7}$ m$^2$ s$^{-1}$. To our knowledge, there are no experimental diffusivity data available for liquid PCMs. Some partial atomic diffusion coefficients are available from *ab initio* computer simulations (*28*) $D_{Ge}$ =4.04 × 10$^{-9}$ m$^2$ s$^{-1}$, and $D_{Te}$ = 4.06 × 10$^{-9}$ m$^2$ s$^{-1}$ at 1000 K for the same composition, which are close to our value ($D \approx$5.7× 10$^{-9}$ m$^2$ s$^{-1}$) at 1003 K.

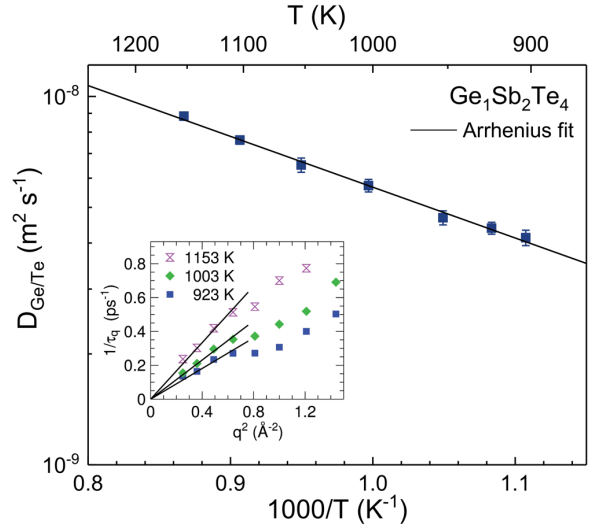

**Figure 2**. **Mean self-diffusion coefficient $D_{Ge/Te}$ of liquid $Ge_1Sb_2Te_4$ derived from incoherent scattering signal at low $q$. Inset:** $q^2$-dependence of the inverse relaxation time $1/\tau_q$. At low-$q$ ($q^2 \leq$0.6 Å$^{-2}$), $1/\tau_q$ follows a $q^2$ -dependence (solid straight lines), as expected from hydrodynamic theory as $q \rightarrow 0$ Å$^{-1}$. For $q^2 > 0.6$ Å$^{-2}$, $1/\tau_q$ deviates from the $q^2$ –dependence due to coherent scattering contributions from a pre-peak of $S(q)$ at ∼ 1 Å$^{-1}$ (see SI-Fig.S3).

## DISCUSSION

According to the SER, the product $(D \cdot \eta)/T$ and, hence, $(D \cdot \tau_\alpha)/T$ should remain constant as a function of $T$ [Eq. (1)]. For the liquid PCM $Ge_1Sb_2Te_4$, this is evidently not the case, as highlighted in Figure 3a. This relation appears to hold at high temperatures ($T \gtrsim$1050 K). However, a marked deviation is observed at 1050 K on approaching $T_m$ (903 K) during cooling, indicating a breakdown of the SER well above the melting point, $T_{SE} = 1.16T_m$. Note that we take the SER in its form of $D \propto (\tau_\alpha/T)^{-1}$. If $G_\infty$ in the Maxwell relation is temperature dependent (as it certainly is for fragile liquids), then the temperature of SER breakdown, in its original form with viscosity, might differ somewhat from the breakdown temperature observed here. In either case it should occur at relaxation times of order 1ps (from Fig. 1 and Fig 3b), far shorter than in any normal liquid where the breakdown only occurs at nanosecond relaxation times.

In Fig.3b, by fitting the data with a fractional SER of the form(*29*),

$$D \propto (\tau_\alpha/T)^{-\xi}, \qquad (2)$$

where $0 < \xi \le 1$, we see that the high-temperature liquid for $T \ge 1050$ K indeed closely follows the SER with an exponent $\xi \approx 0.97 \pm 0.11$, while $\xi \approx 0.60 \pm 0.03$ is obtained for $T \le 1050$ K, indicating a strong deviation from the SER. The latter is related to the decoupling of crystal growth coefficient $U_{kin}$ and viscosity $\eta$ for fragile liquids, which is described by the form $U_{kin} \propto \eta^{-\xi^*}$ given by Ediger et al.(*9*). Indeed, $\xi^*=0.67$ in a similar PCM $Ge_2Sb_2Te_5$, estimated by Orava et al.(*7*) from an empirical correlation with fragility, is close to our $\xi \approx 0.6$.

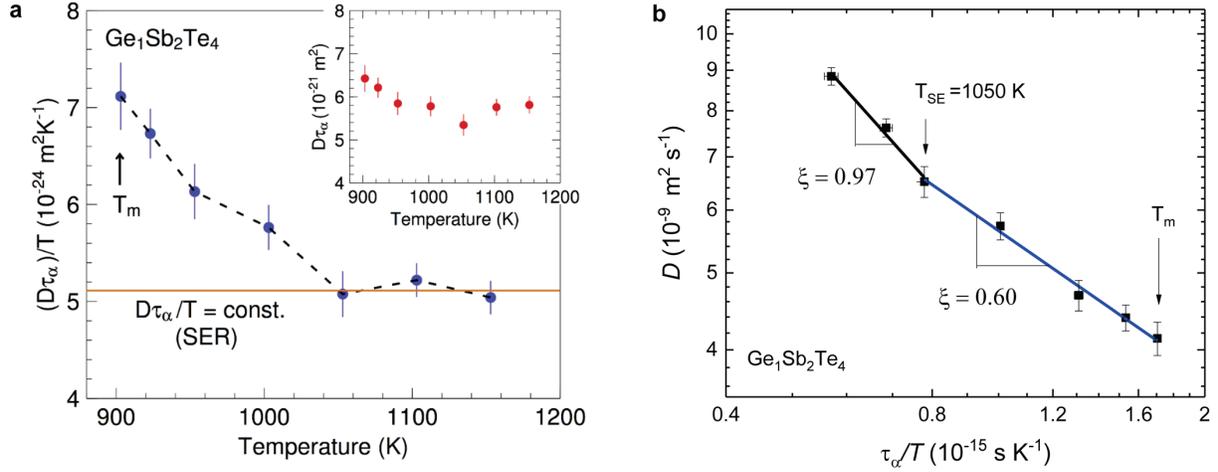

**Figure 3. The breakdown of the Stokes-Einstein Relation in liquid PCMs.** (**a**) During cooling, $(D \cdot \tau_\alpha)/T$ initially has a constant value at $T \ge 1050$ K as expected from the SER, but shows a marked positive deviation from the constant below 1050 K as $T_m =903$ K is approached, indicating a breakdown of the SER at $T_{SE}=1050$ K. **Inset:** The product $D \cdot \tau_\alpha$ does not exhibit a simple proportionality with temperature and clearly takes on a negative slope at lower temperatures. (**b**) The data fitting with the fractional SER indicates a change of slope from $\xi \approx 0.97 \pm 0.11$ to $\xi \approx 0.60 \pm 0.03$ at $T_{SE}=1050$ K, as in Fig 3(a).

The breakdown of the SER at such high temperatures ($>T_m$), and short relaxation times, is an important observation, since (i) the SER has provided the basis for calculating viscosity and fragility from simulated self-diffusion coefficients and/or crystal growth velocities (or vice versa) of PCMs near $T_m$ (*7, 11, 12*), and (ii) it occurs at temperatures where diffusivities are nearly four orders of magnitude higher ($\sim 5 \times 10^{-9}$ m$^2$ s$^{-1}$) than where the SER breakdown is observed in conventional glass formers. For instance, for the typical fragile molecular liquid OTP, the SER remains valid down to the much lower diffusivity $D \approx 1.3 \times 10^{-13}$ m$^2$ s$^{-1}$ (ref. (*30*)) where $\eta \approx 7.7$ Pa·s.

Indeed, the observation in Ge$_1$Sb$_2$Te$_4$ more resembles the case of the "most anomalous liquid" - supercooled water. For bulk water, the recent data of Dehaoui et al.(*31*) showed a crossover in the fractional SER behavior from $\xi \approx 1$ at high temperature to $\xi \approx 0.8$ at low temperature. The breakdown temperature $T_{SE}$ of ~340 K, ~ $1.25T_m$, and the diffusivity $D \approx 5 \times 10^{-9}$ m$^2$ s$^{-1}$ at T$_{SE}$ (where viscosity $\eta \approx 0.4$ mPa s), are comparable to those of Ge$_1$Sb$_2$Te$_4$ (see Fig. S4). Note that the deviation from the SER in water appears well above other known anomalies such as density maximum (277 K), rapid C$_p$ increase, and sharp viscosity rising (below $T_m$). Harris(*32*), with less extensive data, had earlier reported a SER breakdown in water also at $D = 0.59 \times 10^{-9}$ m$^2$ s$^{-1}$ (ref.(*33*)) with crossover to fractional SE behavior with $\xi=0.67$. As is much discussed, the anomalies of water are thought to be related to a LLT, and possibly a nearby, but hidden, second critical point suggested by some computational models(*34*).

To further explore the origin of the SER breakdown in such a high-atomic-mobility state of Ge$_1$Sb$_2$Te$_4$, we turn to the analogous phenomenology found in some relevant computational models. For instance, in the study of the 'two-scale' Jagla ramp potential(*35*) it was found that a breakdown in the SER began at a temperature of 0.6, which is 50% above the temperature of the Widom line crossing at $T_w$. Something similar is also observed in models of water, for instance in ST2 water(*36*), $T_\times$ (which is $T_{SE}$ in our case), was found to be ~10% above $T_w$. Thus, the breakdown itself is not specific to water, but may occur in other systems that possess a liquid-liquid critical point (LLCP). Whether or not the breakdown is due directly to the implied structural heterogeneity is, however, not clear, as the following discussion will emphasize.

As an isobar crosses the 'Widom-line' in the supercritcal region near the LLCP, the populations of high and low density nanodomains reverse, and response functions like $C_p$ maximize(*37*). It is natural then to relate the SER breakdown to the critical-point-related structural fluctuations, and the decoupling of viscosity from diffusion behavior(*37*) that they promote. The problem we see with this appealing association comes from different sources: two from the simulation literature and the other from our own observations, as follows.

In their detailed study of the SER breakdown in ST2 water, Poole and co-authors[29] concluded that both slow and fast subsets of molecules contributed to the deviation and, furthermore, "the fractional behavior is observed across three distinct physical regimes…, indicating an almost complete insensitivity to changes in the average liquid structure." Similar conclusions can be reached by analyzing the diffusivity and $\tau_\alpha$ data provided for the Fermi-Jagla model in ref.[38] (for results, see SI-Fig.S2). In our own work, we have been struck by the fact that the onset temperature for deviation from the SER occurs in a temperature domain where $S(q,t)$ is still perfectly exponential, with no sign, even at temperatures very near the melting point, of the sort of shoulder usually associated with stretching of the exponential and the development of the dynamic heterogeneity (seen, for instance, in ref.[38] (Fermi-Jagla, see SI)). We note that, however, the MD simulation study of the ISF shows the presence of dynamic heterogeneities in the supercooled GeTe[39], which, originating from structural heterogeneities due to chains of Ge-Ge homopolar bonds, may explain the breakdown of the SER in GeTe in the viscous regime below $T_m$ [40]. It is apparently not the case here.

How can we attribute the observed SER breakdown to heterogeneity if, even in the sensitive dynamical properties, we see no sign of heterogeneity? Just as mysterious is the breakdown of SER that we find at temperatures where the relaxation time is so much shorter (ps) than it is in "normal" liquids where SER breakdown occurs e.g. $\tau_\alpha >$ ns for OTP and other glassformers. Is it that the SER is a much more sensitive signaler of impending anomalous character than any of the other signals yet studied? Let us remember that in water the SER breakdown also occurred at picosecond relaxation times.

Given that 'water-like' anomalies such as density maxima, and diverging (or peaked) heat capacities, occur in supercooled Te, Ge, Si[41–44], in $Ge_{15}Te_{85}$ just above the eutectic temperature[45], and in $As_2Te_3$ somewhat above its $T_m$[46], we should be prepared for unusual behavior in the PCMs at lower temperatures[23]. The thermodynamic response function maxima in the above-mentioned chalcogenides are also associated with liquid metal-semiconductor transitions[23]. Indeed, pressure-induced polyamorphic transitions between high- and low-density amorphous states (which are also metallic and semiconducting states) have been found in both $Ge_1Sb_2Te_4$ and $Ge_2Sb_2Te_5$[47–49] and these closely parallel the polyamorphism in amorphous silicon[50] and vitreous ice[51] - except that the latter obviously does not possess a semiconductor-metal transition.

On the basis of all the above, a $P$-$T$ diagram for liquid, and metastable liquid, states of $Ge_1Sb_2Te_4$ is conjectured in Fig. S5. Its analogy to that of water is provided by the inset. Errington and Debenedetti[52] showed that a "Russian doll" of nested kinetic and thermodynamic anomalies exists in water and the same has since been found for other water-like systems (e.g. Si and $SiO_2$(*53, 54*)). The anomaly persisting to the highest temperature is the structural characteristic related to openness of structure -- or tetrahedrality in the case of water. The figure suggests that, as theorized for water, a metastable critical point might exist not far from ambient pressure in liquid PCMs and not only give rise to the SER breakdown above $T_m$, but also facilitate crystallization processes below $T_m$, as has previously been argued for the cases of globular proteins, and some colloidal and Lennard-Jones fluid systems which possess a LLCP(*55–58*). Indeed, Tanaka's two-order-parameter model already predicts that a "V-shape" *P-T* phase diagram, as is for this PCM case, is directly related to thermodynamic and dynamic anomalies similar to those of water(*59*). The fact that we observe the same sort of breakdown for PCMs lends credence to the suggested phase diagram even though a proper understanding of the breakdown remains elusive. The proposed scenario of a LLT is supported by the evidence of a fragile-strong crossover/transition found recently in a similar composition $Ge_2Sb_2Te_5$ below $T_m$, manifesting as a continuous crossover argued by Chen et al.(*15*), and as a singular temperature (792 K) argued by Flores-Ruiz et al.(*60*) by data fitting.

What do the above considerations (assuming their validity) imply for the deeper understanding of PCM function and related future research in this area? Space limitations compel brevity, so we refer to previous discussions of the felicitous effect of a fragile-to-strong transition occurring shortly below $T_m$, by speeding crystallization on the high temperature side of the peak and arresting the dynamics by a fragile-to-strong transition on passing the peak(*23, 24, 45, 61*). Urgent for improvement of this understanding is the elucidation of the character of the glasses produced by hyperquenching, and finding their relation to alternative polyamorphs mentioned above. If a first-order transition is involved, then it should be possible to trap the microdomains that coexist transiently during the hyperquench and analyze them by TEM coupled with EDX and related techniques. For continuous transitions, density studies of the glassy phases, trapped by quenching at different rates, should be diagnostic. Glasses made using different sputter schedules, should offer an alternative approach. Acquired data would permit a mapping out of the features of the potential energy hypersurface(*62*) ("energy landscape"), on which the system is being trapped. Is it a single landscape or is there a discontinuous gap (first-order transition)(*63*)?

It is important that the phenomena described in this work are not confused with recent reports of SER breakdown above "$T_m$" (a ternary eutectic temperature) in certain glass-forming metallic mixtures, such as ZrCuAl, where the species Cu decouples from the Zr-Al matrix(*64, 65*). The latter phenomenon is more closely related to the case of Cu in amorphous silicon, where $D_{Cu}$ can be four orders of magnitude greater than that of the host (Si) atoms(*66*). A related phenomenon, also quite different from our conjecture, is the superionicity of Cu (or Ag) cations in many superionic glassformers, where the mobile ion decoupling is observed well above any liquidus temperature, (and also above $2T_g$), and the Stokes-Einstein discrepancy at $T_g$ can reach 11 orders of magnitude(*67*)).

## SUMMARY

We have performed neutron scattering studies of diffusion and relaxation times in the PCM $Ge_1Sb_2Te_4$, and identified a breakdown in the SER well above the $T_m$ - which lies in a relaxation time domain $10^4$ times shorter than in normal liquids. We link this to the behavior observed in liquid silicon, germanium, and water, where it is seen as a consequence of a submerged liquid-liquid transition which provokes facile crystallization and fragile-to-strong transitions when ultrafast cooling preserves the liquid state. The exploration of PCMs' anomalous liquid-state behavior will be an essential step towards understanding the fast phase switching behavior in this class of material.

## MATERIALS AND METHODS

### Sample preparation

$Ge_1Sb_2Te_4$ was prepared using the Ge, Sb, and Te elements with purities ranging from 99.999 to 99.9999 at. %. The elements were sealed under vacuum ($10^{-6}$ mbar) in a fused quartz tube with internal diameter of 5 mm and synthesized in a rocking furnace for homogenization at 900°C for 15 hours.

### QENS measurements and data analysis

The sample, sealed in the fused quartz tube, was loaded into a thin-walled $Al_2O_3$ container for the QENS measurements, which were carried out at the time-of-flight spectrometer TOFTOF at the Heinz Maier-Leibnitz neutron source (FRM II) in Munich(*68*, *69*). Two incident neutron wavelengths $\lambda_i = 4.4$ and 7 Å were used to obtain a broad q and energy transfer range along with a high resolution of about 90 μeV (full width at half maximum).

Spectra were collected as a function of temperature in a high-vacuum high-temperature Nb-furnace. Raw time-of-flight data were normalized to a vanadium standard and interpolated to constant q to obtain the dynamic structure factor $S(q, \omega)$ using the FRIDA-1 software [See http://sourceforge.net/projects/frida/ for source code.] All spectra were found to be well described by a model composed of the quasi-elastic scattering from the alloy melt and a flat background to approximate the processes too fast to be accurately measured by the spectrometer. The $S(q,\omega)$ obtained in the measurements where $\lambda_i = 7$ Å were additionally modeled to include the elastic scattering from the container. In general, the model $S(q,\omega)$ reads:

$$S(q,\omega) = R(q,\omega) \otimes N \left[ A_0 \delta(\omega) + (1 - A_0) L(q,\omega) \right] + b(q,\omega),$$

where $R(q,\omega)$ is the instrumental resolution function, $N$ is a normalization factor, $A_0$ is the magnitude of the elastic scattering and $b(q,\omega)$ is a constant but $q$-dependent, background. The symbol $\otimes$ denotes a numerical convolution. The quasi-elastic scattering was found to be best described with a single Lorentzian of the form (see SI-Fig.S1),

$$L(q,\omega) = \frac{1}{\pi} \frac{\Gamma(q)}{(\hbar\omega)^2 + \Gamma(q)^2},$$

where $\Gamma$ is the half-width at half-maximum. Below $q^2 \sim 0.6$ Å$^{-2}$ the incoherent scattering from Ge and Te dominates and the coherent contributions from thermal diffusion (Rayleigh line) and acoustic modes are effectively contained in the flat background of the observed quasi-elastic spectra(70). A mean Ge/Te self-diffusion coefficient was determined via

$$D_{Ge/Te} = \frac{\Gamma(q)}{\hbar q^2}.$$

An analysis was carried out also in the time domain first by obtaining the intermediate scattering function $S(q,t)$ (or density correlation function) via cosine Fourier transform of the measured $S(q,\omega)$ and normalizing to the instrumental resolution function $R(q,t)$. In general, the data were then fitted with a simple exponential decay as

$$S(q,t)/S(q,0) = f(q) \exp[-t/\tau(q)] + c,$$

where $f(q)$ is the amplitude, $\tau$ is the structural relaxation time and the constant $c$ is an offset that takes care of any remaining elastic scattering. It should be noted that this is in line with the model used for $S(q,\omega)$, as the Fourier transform of a Lorentzian is indeed a simple exponential. In order to ensure consistency of the analyses in both energy transfer and time domain, we restricted the fitting range in the energy transfer domain to [-1,1] meV and in the time domain to the data points after 0.65 ps. At higher energy transfers and shorter times, the spectra are dominated by phononic vibrations and fast relaxation processes. The self-diffusion coefficient was obtained from the time domain analysis via

$$D_{Ge/Te} = \tau(q)^{-1} q^{-2}.$$

The values of $D_{Ge/Te}$ reported in the manuscript represent an average of the values obtained in both analyses.

**Acknowledgements**

The authors acknowledge the beamtime at the TOFTOF instrument operated by FRM II at the Heinz Maier-Leibnitz Zentrum (MLZ), Garching, Germany, and financial support provided by FRM II to perform the QENS measurements. S.W. acknowledges the support from Alexander von Humboldt Foundation Feodor Lynen Postdoctoral Research Fellowship, Place-to-be RWTH Start-Up fund, and the DFG within SFB917. P.L. acknowledges financial support from NSF-EFRI award No. 1640860. C.A.A. acknowledges support from National Science Foundation Research Grant No. CHE-1213265.

**SUPPLEMENTARY MATERIALS**

**Supplemental Figures**

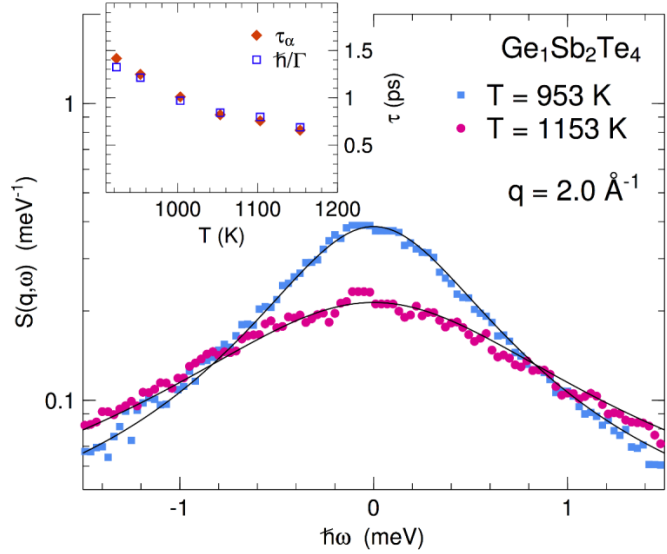

**Figure S1. The dynamic structure factor S(q,ω) in the energy transfer domain (ℏω) obtained from quasi-elastic neutron scattering**. The S(q,ω) was found to be best described with a single Lorentzian of the form

$$L(q,\omega) = \frac{1}{\pi} \frac{\Gamma(q)}{(\hbar\omega)^2 + \Gamma(q)^2},$$

where $\Gamma$ is the half-width at half-maximum. The inset shows that the $\hbar/\Gamma$ values obtained from the energy domain are in good agreement with the $\tau_\alpha$ from the analysis carried out in the time domain (see Fig.1a and text), where the intermediate scattering function S(q,t) is obtained via cosine Fourier transform of the measured S(q,ω) and normalized to the instrumental resolution function R(q,t).

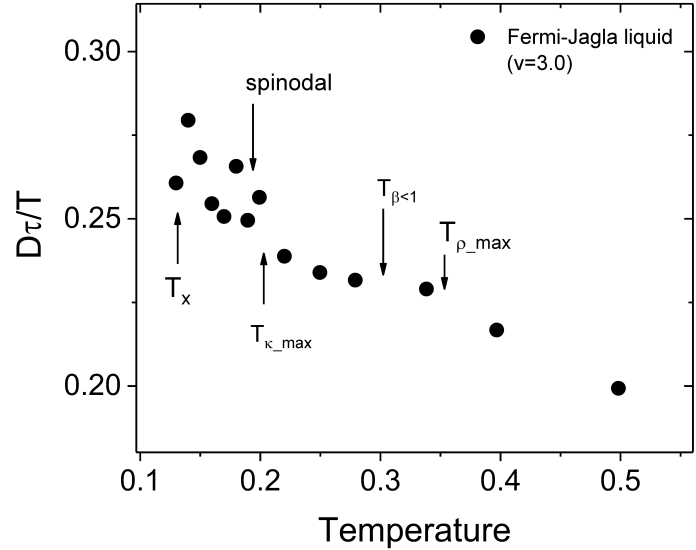

**Figure S2. The $D\tau_\alpha/T$ as a function of temperature for the Fermi-Jagla liquid at a volume v=3.0 (close to the liquid-liquid critical point volume $v_c \approx 2.9$).** Data are adapted from Sun et al. (J. Chem. Phys. 146, 2017). The model liquid exhibits anomalous properties with decreasing temperature. The arrows indicate the temperature of density maximum $T_{\rho\_max} \approx 0.35$, the onset temperature of the non-exponential decay, $T_\beta \approx 0.3$, the temperature of compressibility maximum, $T_{\kappa\_max} \approx 0.2$, the spinodal line for the liquid-liquid coexistence (close to the critical temperature $T_c \approx 0.18$), and the crystallization temperature $T_x \approx 0.13$. We note that the data of $D\tau_\alpha/T$ show a negative slope as a function of temperature, indicating a violation of the SER within the entire accessible temperature range from 0.5 to $T_x$. This suggested that the onset of the SER breakdown is likely to be at a higher temperature T>0.5, out of the accessible window. By contrast, the decay of the intermediate scattering functions only become non-exponential at $T_\beta \approx 0.3$ which is even lower than $T_{\rho\_max}$.

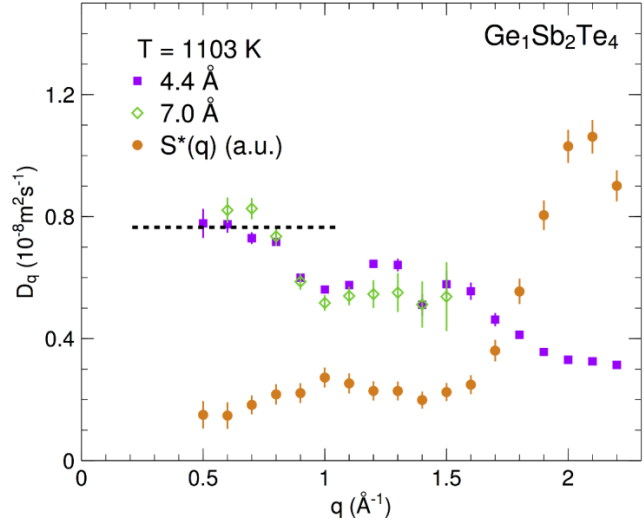

**Figure S3.** The '*effective*' diffusion coefficient $D_q$ as a function of $q$ derived at a given temperature of 1103 K using the relation $D_q = 1/(\tau_q q^2)$. The $D_q$ measured at both incoming wavelengths of 4.4 Å and 7 Å are generally in good agreement with each other. The dashed line gives the long-range self-diffusion coefficient in the low-q limit. Integrating the measured S(q,ω) in an energy transfer range from -0.3 to 0.3 meV yields a quantity $S^*(q)$ similar to the true static structure factor S(q). The reduction in $D_q$ near the main $S^*(q)$ peak at ~2 Å$^{-1}$ was predicted by de Gennes (De Gennes, Physica 25, 825–839 (1959)), and describes the slowing down of the microscopic dynamics of a liquid in the presence of structural ordering. This feature is also reproduced in the vicinity of the so-called pre-peak in $S^*(q)$ at $q \approx 1$ Å$^{-1}$, which is indicative of a distinct MRO occurring in liquid Ge$_1$Sb$_2$Te$_4$. Note that, as $q$ approaches the pre-peak from lower values, $1/\tau_q$ deviates from the $q^2$–dependence, as shown in Fig.2 Inset.

**Comparisons of water and GST deviations from the SER**

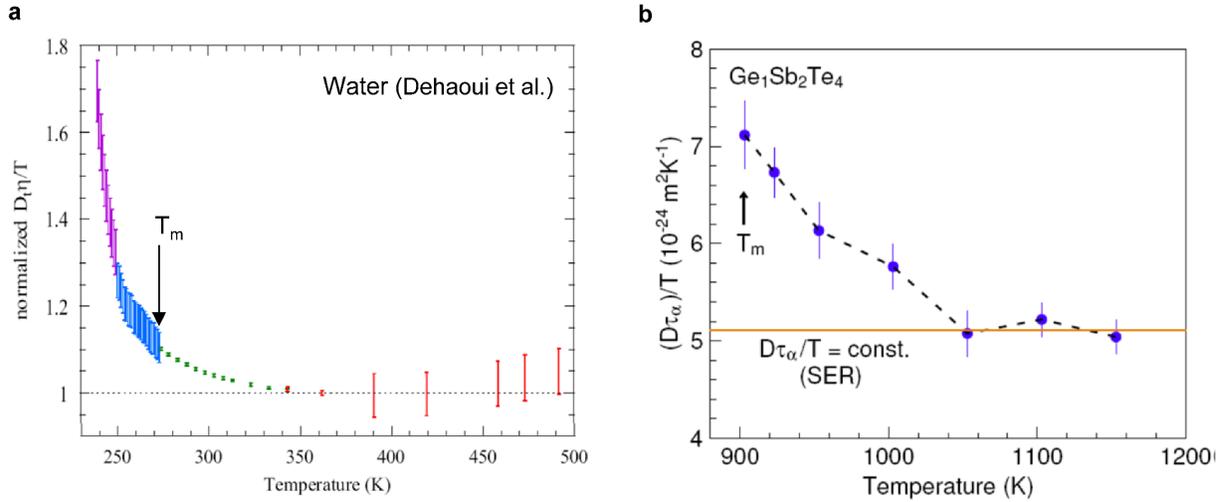

**Figure S4.** Comparison of the SER breakdown in water (a) [adapted from Dehaoui et al.(*31*)] and liquid Ge$_1$Sb$_2$Te$_4$ (b). The test of the SER in water uses a combined dataset of viscosity η and translational diffusivity D$_t$ with small uncertainties that covers a full temperature range. The sources of the viscosity and translational diffusivity data are given in Table S3 in SI of [31].

Note that the recent study of the SER in water indicates that the liquid deviates from the SER already at around 340 K, which is ~1.25T$_m$ and higher than that reported by Harris (258 K). Thus, the breakdown of the SER appears to be the very sensitive early sign of the thermodynamic anomalies that occur at lower temperatures (e.g. density maximum, C$_p$ increase, viscosity rising etc.).

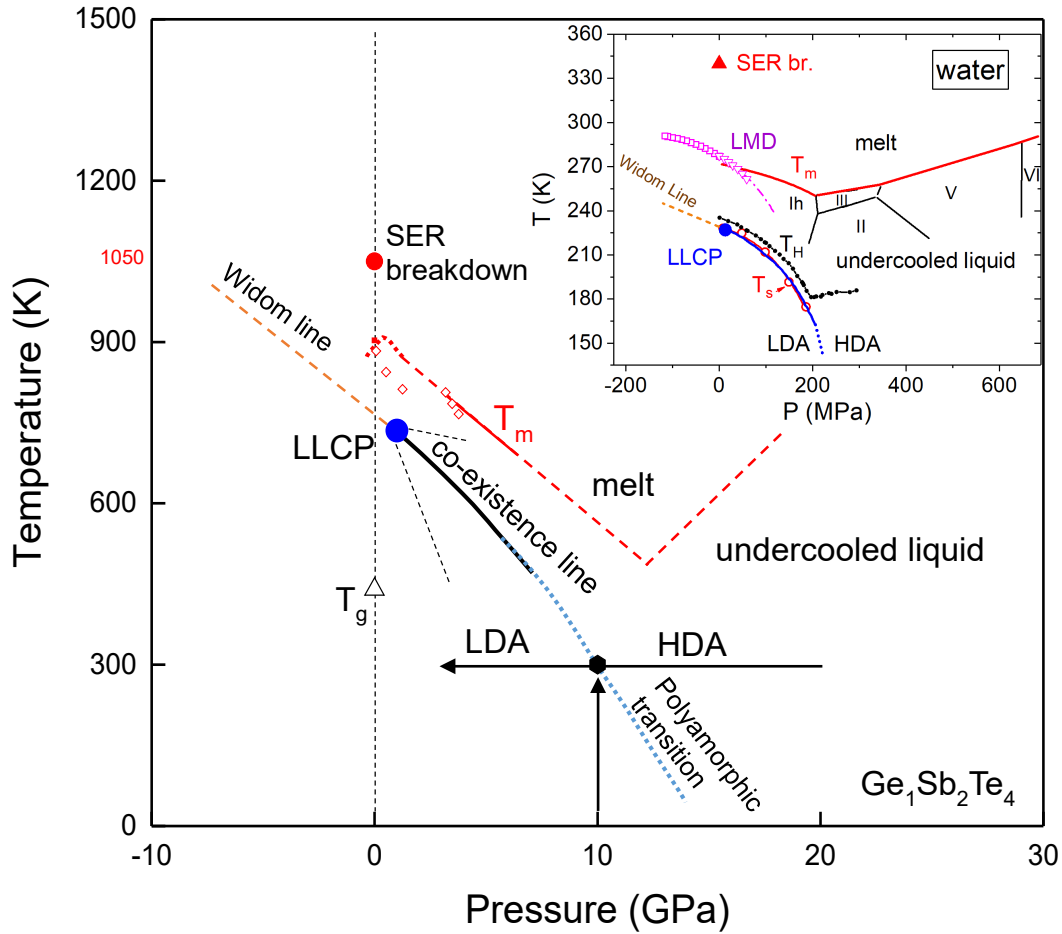

**Figure S5. Pressure-temperature (P-T) metastable liquid phase diagram conjectured for Ge₁Sb₂Te₄.** A liquid-liquid critical point (LLCP) is placed ~30 % below the point of the SER breakdown $T_{SE}$=1050 K, or ~20 % below $T_m$. In the pressure domain, the LLCP is assumed to lie at slightly positive pressure, and under the local $T_m$-curve maximum. The latter can be justified from the fact that a positive volume change upon melting (i.e. positive Clapeyron slope) is reported(*21*, *71*) at ambient pressure while the high-pressure V-shape $T_m$-curve (and open diamonds) accords with the work of Sen and coworkers(*72*, *73*). Therefore, the ambient pressure isobar crosses a Widom line (dashed bold line) with a LLCP nearby. At more positive pressures, the speculated liquid-liquid coexistence line has a negative slope to reconcile the negative Clapeyron-slope of the pressure-induced polyamorphic transition (blue dotted line) between high- and low-density amorphous phases (HDA and LDA) of Ge₁Sb₂Te₄ at ~10 GPa reported by Kalkan et al.(*47*). **Inset:** *P-T* phase diagram of supercooled water. The phase boundaries of melt and ices are taken from ref.(*74*, *75*). The line of maximum densities (LMD) consists of data from an equation of state (open squares) based on sound velocities at negative pressure(*76*) and data at positive pressure(*34*) (open triangles), which are supported by extrapolations of the Holten-

Anisimov two-state equation of state (dash-dotted line) up to its stated reliability limit of -100 MPa(*77*). The SER breaks down in bulk water at ~340 K (solid triangle) well above $T_m$ according to Dehaoui et al.(*31*). $T_H$ is the limit of the homogeneous nucleation temperatures(*74*). Red open circles and line ($T_s$) indicate the singular temperatures(*74*), coinciding with the hypothesized LLCP and the phase boundary of HDA/LDA water(*77*).